\documentstyle[12pt]{article}
\newcommand{\be}{\begin{equation}}
\newcommand{\ee}{\end{equation}}
\newcommand{\bea}{\begin{eqnarray}}
\newcommand{\eea}{\end{eqnarray}}
\newcommand{\ba}{\begin{array}}
\newcommand{\ea}{\end{array}}

\newcommand{\norsl}{\normalsize\sl}
\newcommand{\norsc}{\normalsize\sc}
\begin{document}
\input epsf.sty

\begin{titlepage}

\title{
Target Mass Corrections to \\
QCD Bjorken Sum Rule for \\
Nucleon Spin Structure Functions\thanks{Talk presented by T.~Uematsu 
at the YITP Workshop \lq\lq 
From Hadronic Matter to Quark Matter~\rq\rq, Kyoto, Japan,
Oct.~30 - Nov.~1, 1994. To appear in the proceedings.} 
\\
\quad \\
\quad \\
}
\author{  
\norsc  Hiroyuki KAWAMURA and 
         Tsuneo UEMATSU
\quad \\
\norsl  Department of Fundamental Sciences\\
\norsl  FIHS, Kyoto University\\
\norsl  Kyoto 606-01, JAPAN}

\date{}
\maketitle

\vspace{3cm}

\begin{abstract}
{\normalsize

We discuss the possible target mass corrections in the QCD analysis 
of nucleon's spin-dependent structure functions measured in the 
polarized deep-inelastic leptoproduction.
The target mass correction for the QCD Bjorken sum rule is obtained
from the Nachtmann moment and its magnitude is estimated employing 
positivity bound as well as the experimental data for the asymmetry 
parameters. We also study the uncertainty due to target mass effects in
determining the QCD effective coupling constant $\alpha_s(Q^2)$ from the 
Bjorken sum rule. The target mass effect for the Ellis-Jaffe sum rule
is also briefly discussed. 
}
\end{abstract}

\begin{picture}(5,2)(-335,-670)
\put(2.3,-125){KUCP-76}
\put(2.3,-140){January 1995}
\end{picture}

\vspace{0.5cm}

\thispagestyle{empty}
\end{titlepage}
\setcounter{page}{1}
\baselineskip 20pt


\vspace{0.3cm}
\leftline{\large\bf 1. Introduction}
\vspace{0.3cm}

I would like to talk about possible target mass corrections in the 
QCD analysis of spin-dependent structure functions which can be measured 
by the deep inelastic scattering of polarized leptons on polarized nucleon 
targets \cite{SLAC,EMC}. We especially investigate those for the QCD Bjorken 
sum rule.

As discussed in the previous talks by Dr. Kodaira and Dr. Sloan, the
Bjorken sum rule with QCD corrections is given by
\be
\int_0^1 dx \bigl [ g_1^p(x,Q^2)-g_1^n(x,Q^2) \bigr ]
=\frac{1}{6} \frac{G_A}{G_V} \bigl [ 1-\frac{\alpha_s(Q^2)}{\pi}+
{\cal O}(\alpha_s^2) \bigr ],
\label{bj}
\ee
where $g_1^p(x,Q^2)$ and $g_1^n(x,Q^2)$ are the spin structure function $g_1$
of proton and neutron, respectively, with $x$ and $Q^2$ being the Bjorken 
variable and the virtual photon mass squared. On the right-hand side, 
$G_A/G_V \equiv g_A$ is the ratio of the axial-vector to vector coupling 
constants.
The QCD correction of the order of $\alpha_s(Q^2)$ was first obtained some 
years ago based on operator product expansion (OPE) and renormalization 
group (RG) method in ref. \cite{KMMSU,KMSU}. The higher order corrections were
calculated in refs. \cite{GL,LV,LTV,KS}. 

Now recent experiments on the spin structure function $g_1(x,Q^2)$ for the
deuteron, $^3$He and the proton target at 
CERN and SLAC \cite{SMCD,E142,SMCP,E143} together 
with EMC data have provided us with the data for testing the Bjorken sum
rule \cite{Bj}as well as Ellis-Jaffe sum rule \cite{EJ}. 
In order to confront the QCD prediction with the experimental data at
low $Q^2$ where the QCD corrections are significant, we have to take into
account the corrections due to the mass of the target nucleon, which we
denote by $M$. In this $Q^2$ region, we cannot neglect the order $M^2/Q^2$ 
term, which consists
of higher-twist effects as well as target mass effects.  Here we shall not 
discuss the higher-twist effect which was mentioned by Dr. Kodaira and will 
be discussed by Dr. Mueller this afternoon.  Here we confine ourselves to the
target mass effects.  Here we observe that 1) The target mass
effects are calculable without any ambiguity ; 2) The infinite power series 
in $M^2/Q^2$ can actually be summed up into a closed analytic form.   

In the framework of OPE, the target mass effects of structure functions
can be evaluated by taking account of trace terms of composite operators 
to have a definite spin projection. This amounts to replace the ordinary
moments of the structure functions by the Nachtmann moments \cite{N}.
The Nachtmann moments for spin structure functions were obtained 
in refs.\cite{WANDZ,MU}.
In ref. \cite{WANDZ} the Nachtmann moments were given as an infinite power
series, while in ref. \cite{MU} they were obtained as a closed analytic form.
 
\vspace{0.3cm}
\leftline{\large\bf 2. OPE and Target Mass Effects}
\vspace{0.3cm}

The anti-symmetric part of the virtual Compton amplitude can be written in
the OPE \cite{HM}: 
\bea
&&T_{\mu\nu}(p,q,s)^{[A]}
\simeq
-i\varepsilon_{\mu\nu\lambda\sigma}q^{\lambda}\hspace{-0.2cm}
\sum_{n=1,3,\cdots}
{\Bigl({2\over{Q^2}}\Bigr)}^n q_{\mu_1}\cdots q_{\mu_{n-1}}
E_{1}^n(Q^2,g) 
\ \langle p,s|R_{1}^{\sigma\mu_1\cdots\mu_{n-1}}
|p,s\rangle \nonumber\\
&&\quad -i(\varepsilon_{\mu\rho\lambda\sigma}q_{\nu}q^{\rho}-
\varepsilon_{\nu\rho\lambda\sigma}q_{\mu}q^{\rho}-q^2
\varepsilon_{\mu\nu\lambda\sigma}) 
\hspace{-0.2cm}\sum_{n=3,5,\cdots}
{{n-1}\over n}{\Bigl({2\over{Q^2}}\Bigr)}^n
q_{\mu_1}\cdots q_{\mu_{n-2}}E_{2}^n(Q^2,g) \nonumber\\
&&\hspace{9cm}\times \langle p,s|R_{2}^{\lambda\sigma\mu_1\cdots\mu_{n-2}}
|p,s\rangle,
\label{OPE}
\eea
where the nucleon matrix element of the twist-2 operators 
$R_1^n$ is given by
\bea
\langle p,s|R_1^{\sigma\mu_1\cdots\mu_{n-1}}
|p,s\rangle
&=&-a_n[\{s^{\sigma}p^{\mu_1}\cdots p^{\mu_{n-1}}\} -\mbox{trace terms}]
\nonumber\\
 &\equiv& -a_n\{s^{\sigma}p^{\mu_1}\cdots p^{\mu_{n-1}}\}_n,
\label{r1}
\eea
which is totally symmetric in Lorentz indices $\sigma, \mu_1, \cdots, 
\mu_{n-1}$ and traceless. ($\{ \ \}$ denotes symmetrization.) 
While for the twist-3 operator $R_2^n$, the matrix element is given by
\bea
\langle p,s|R_2^{\lambda\sigma\mu_1\cdots\mu_{n-2}}
|p,s\rangle
&=&-d_n[\frac{1}{2}(s^{\lambda}p^{\sigma}-s^{\sigma}p^{\lambda})
p^{\mu_1}\cdots p^{\mu_{n-2}} -\mbox{trace terms}]
\nonumber\\
 &\equiv& -d_n\{s^{\lambda}p^{\sigma}p^{\mu_1}\cdots p^{\mu_{n-2}}
\}_{M_n},
\label{r2}
\eea
which is symmetric in $\sigma\mu_1\cdots\mu_{n-2}$ and anti-symmetric in
$\lambda\sigma$ and also traceless in its Lorentz indices.

Taking account of trace terms we can project out the contribution from
a definite spin as follows.  By taking a contraction of the tensor appearing
in eq.(\ref{r1}) with $q_{\mu_1}\cdots q_{\mu_{n-1}}$ we get
\bea
q_{\mu_1}\hspace{-0.2cm}&\cdots&\hspace{-0.2cm} q_{\mu_{n-1}}
\ [\{s^{\sigma}p^{\mu_1}\cdots p^{\mu_{n-1}}\} -(\mbox{trace terms})]
\nonumber\\
&=&\hspace{-0.2cm}\frac{1}{n^2}[s^{\sigma}a^{n-1}C_{n-1}^{(2)}(\eta)
+q^{\sigma}\frac{q\cdot s}{Q^2}a^{n-1}\times 4C_{n-3}^{(3)}(\eta) 
+p^{\sigma}q\cdot s a^{n-2}\times 2C_{n-2}^{(3)}(\eta)],
\label{gegen}
\eea
where $C_n^{(m)}(\eta)$ is a Gegenbauer polynomial with
\be
\eta=i\nu/Q, \ \nu=p\cdot q/M, \ a=-\textstyle{\frac{1}{2}}iMQ.
\ee
Using the orthogonality property of Gegenbauer polynomials, one
can project out the contribution from operators with a definite spin.
The closed analytic forms for the Nachtmann moments are given as
\cite{MU}:
\bea
&&M_1^n(Q^2)\equiv \textstyle{\frac{1}{2}}a_nE_1^n(Q^2,g) \nonumber\\
&&\quad=\int_0^1\frac{dx}{x^2}\xi^{n+1}\Bigl [ \bigl\{ \frac{x}{\xi}-
\frac{n^2}{(n+2)^2}\ \frac{Mx}{Q}\ \frac{M\xi}{Q}\bigr\}
g_1(x,Q^2)-\frac{4n}{n+2}\frac{M^2x^2}{Q^2}g_2(x,Q^2) \Bigr ],
\nonumber\\
&& \hspace{10cm}\quad (n=1,3,\cdots)
\label{m1}\\
&&M_2^n(Q^2)\equiv \textstyle{\frac{1}{2}}d_nE_2^n(Q^2,g) \nonumber\\
&&\quad=\int_0^1\frac{dx}{x^2}\xi^{n+1}\Bigl [ \frac{x}{\xi}
g_1(x,Q^2)+\bigl\{\frac{n}{n-1}\frac{x^2}{\xi^2}
-\frac{n}{n+1}\frac{M^2x^2}{Q^2}\bigr\}g_2(x,Q^2) \Bigr ], \nonumber\\
&& \hspace{10cm}\quad (n=3,5,\cdots)
\label{m2}
\eea
where $\xi$ is a variable given by \cite{GP}:
\be
\xi\equiv \frac{2x}{1+\sqrt{1+4M^2x^2/Q^2}}.
\ee
Taking the difference between the first moment for the proton target and that 
for the neutron in eq.(\ref{m1}), we can arrive at the QCD Bjorken sum rule 
with target mass correction:
\bea
& &{1\over 9}\int_0^1 dx {{\xi^2}\over{x^2}} \Bigl [
5+ 4\sqrt{1+{{4M^2x^2}\over{Q^2}}} \Bigr ] 
\Bigl [ g_1^p(x,Q^2)-g_1^n(x,Q^2)\Bigr ] \nonumber\\
& &-{4\over 3}\int_0^1 dx {{\xi^2}\over{x^2}}{{M^2x^2}\over{Q^2}}
\Bigl [ g_2^p(x,Q^2)-g_2^n(x,Q^2)\Bigr ]
={1\over 6}{G_A\over G_V}\Bigl [1-{{\alpha_s(Q^2)}\over \pi}+O(\alpha_s^2)
\Bigr ] .
\label{nacht}
\eea
 
Note that in the presence of target mass correction, the other spin structure
function $g_2^{p,n}(x,Q^2)$ also comes into play in the Bjorken sum rule.
Here we emphasize that the target mass correction treated through the above 
procedure is not mere a power correction but given as a closed analytic form.
It should also be noted that target mass corrections considered as the 
expansion in powers of $M^2/Q^2$ is not valid when $M^2/Q^2$ is of order 
unity \cite{SV,BBK}.

Our result can be compared with the target mass correction as a 
power correction discussed in the literatures.
Expanding our Nachtmann moment in powers of $M^2/Q^2$ we get
\bea
\int_0^1dxg_1^{p-n}(x,Q^2)&=&\frac{1}{6}\frac{G_A}{G_V}(1-\frac{\alpha_s}{\pi}
+\cdots ) \nonumber\\
&+&\frac{10}{9}\frac{M^2}{Q^2}\int_0^1dxx^2g_1^{p-n}(x,Q^2)
+\frac{12}{9}\frac{M^2}{Q^2}\int_0^1dxx^2g_2^{p-n}(x,Q^2),
\label{pnacht}
\eea
which coincides with the result given by Balitsky-Braun-Kolesnichenko in 
ref.\cite{BBK} up to the contribution from the twist-4 operator to the
order of $1/Q^2$.

Now, the difference between the left-hand side of (\ref{nacht}) and 
that of (\ref{bj}) leads to the target mass correction $\Delta\Gamma$:
\bea
\Delta\Gamma &=&\int_0^1dx \bigl \{\frac{5}{9}\frac{\xi^2}{x^2}+
\frac{4}{9}\frac{\xi^2}{x^2}\sqrt{1+\frac{4M^2x^2}{Q^2}}-1\bigr\}
\times \Bigl [ g_1^p(x,Q^2) - g_1^n(x,Q^2) \Bigr ] \nonumber\\
& &-{4\over 3}\int_0^1 dx {{\xi^2}\over{x^2}}{{M^2x^2}\over{Q^2}}
\Bigl [ g_2^p(x,Q^2)-g_2^n(x,Q^2)\Bigr ].
\eea

\vspace{0.3cm}
\leftline{\large\bf 3. Estimation of the Target Mass Effects}
\vspace{0.3cm}

Let us now study the size of the target mass correction $\Delta\Gamma$ to
the Bjorken sum rule.
First we note that the spin structure functions $g_1$ and $g_2$ are
written in terms of virtual photon asymmetry parameters $A_1$ and 
$A_2$, which are measured at the experiments, together with
the unpolarized structure function, $F_2(x,Q^2)$, 
and the ratio of the longitudinal to transverse
virtual photon cross sections, $R=\sigma_L/\sigma_T$ \cite{SLAC,EMC}.
 We shall estimate the upper bound for the target mass correction of
$\Delta\Gamma$, which we denote by $\Delta\Gamma_{u.b.}$ 
(i.e. $|\Delta\Gamma| \leq \Delta\Gamma_{u.b.}$) in a variety of methods. 

For the first analysis (Analysis I), we apply the positivity bound for the 
asymmetry parameters \cite{DDR}:
\be
|A_1| \leq 1, \qquad |A_2| \leq \sqrt{R}.
\label{a1a2}
\ee
We use the parametrization for $R$ taken from the global fit of the SLAC data
\cite{RSLAC} and the NMC parametrization for $F_2(x,Q^2)$ \cite{NMC}.
In Fig.1 we have plotted the upper bound, $\Delta\Gamma_{u.b.}$,
as a function of $Q^2$ for Analysis I by a solid line.  
Here the error of the upper bounds of $\Delta\Gamma$
due to the parametrizations $R$ and $F_2$ is typically around 10 $\%$.

In our second analysis (Analysis II), we employ the experimetal data on 
spin asymmetry $A_1$ and positivity bound for $A_2$ to improve the upper 
bound. We take the data on $A_1^p$ from SMC data \cite{SMCP}
together with EMC data \cite{EMC} and those for $A_1^d$ from 
SMC group \cite{SMCD} to extract 
$A_1^n$, for which we can also use the E142 data \cite{E142}. 
For this case, the upper bound is shown in Fig.1 by the short-dashed 
line, which is located slightly lower than $\Delta\Gamma_{u.b.}$ for 
Analysis I. 
When we decompose the $\Delta\Gamma_{u.b.}$ into two parts, $\Delta\Gamma_1$ 
and $\Delta\Gamma_2$, which are the contributions from $A_1$ and $A_2$, 
respectively, it turns out that $\Delta\Gamma_2$ is much larger than 
$\Delta\Gamma_1$. The value of $\Delta\Gamma_1$ turns out be less than 10 
$\%$ of $\Delta\Gamma_2$. 

The third analysis (Analysis III) uses the recently measured $A_2^p$ by the 
SMC group \cite{SMCA2} in addition to the same data for $A_1^{p,n}$ together 
with the positivity bound for $A_2^n$ as in Analysis II. 
We have also plotted the upper bound for Analysis III in Fig.1 by the 
long-dashed line. Here we took the data
on $A_2^p$ obtained by SMC group at the first measurement of transverse 
asymmetries \cite{SMCA2}, where the number of data points are still four and 
the relative error bars are not so small. The $A_2^p$ measured is much 
smaller than the positivity bound. 
If the $A_2$ for the neutron is also small as mentioned in ref.\cite{E142}, 
the $\Delta\Gamma_{u.b.}$ becomes very small.

Finally we briefly comment on uncertainty due to target mass effects in 
determining the QCD coupling constant from Bjorken sum rule which has 
recently been discussed by Ellis and Karliner \cite{EK2}.
From the QCD corrections up to ${\cal O}(\alpha_s^4)$ \cite{LV,LTV,KS}
they obtained the value $\alpha_s(Q^2=2.5\mbox{GeV}^2)=0.375_{-0.081}
^{+0.062}$\cite{EK2}, by using the known $g_A={G_A}/{G_V}$ ratio 
and taking the value $\Gamma(Q^2=2.5\mbox{GeV}^2)$ $=$ 
0.161$\pm$0.007$\pm$0.015, in their analysis of E142 and E143 data \cite{EK2}.
The $Q^2=2.5\mbox{GeV}^2$ is the averaged value of the mean $Q^2$ of the 
E142 data ($<Q^2>\simeq 2\mbox{GeV}^2$) and the E143 data 
($<Q^2>\simeq 3\mbox{GeV}^2$).
Here we shall not take into account the higher-twist effects which are
considered to be rather small as claimed in refs. \cite{EK2}.

The uncertainty in $\Gamma$ due to target mass effects gives rise to that 
for the QCD coupling constant $\alpha_s(Q^2=2.5\mbox{GeV}^2)$.
Namely, $\Delta\Gamma_{u.b.}(Q^2=2.5\mbox{GeV}^2)$=0.029, 0.027 and
0.011 for Analyses I, II and III, respectively, we get
the ambiguities for $\alpha_s$
\bea
0.213 &\leq& \alpha_s(Q^2=2.5\mbox{GeV}^2) \leq 0.474 \quad 
(\mbox{Analysis I}), \nonumber\\
0.228 &\leq& \alpha_s(Q^2=2.5\mbox{GeV}^2) \leq 0.469 \quad
(\mbox{Analysis II}), \nonumber\\
0.315 &\leq& \alpha_s(Q^2=2.5\mbox{GeV}^2) \leq 0.424 \quad 
(\mbox{Analysis III}).
\eea

\vspace{0.3cm}
\leftline{\large\bf 4. Conclusion}
\vspace{0.3cm}

In this talk we have examined the possible target mass corrections to the 
Bjorken sum rule using positivity bound and experimental data on asymmetry
paprmeters. We have found that at relatively small $Q^2$ where the QCD
effect is significant, the target mass effects are also non-negligible.
We found that to test the target mass correction precisely, we need 
accurate data for $A_2(x,Q^2)$.  
In determining the QCD coupling constant $\alpha_s$ from the Bjorken sum
rule, there appears uncertainty due to target mass effects. 
This uncertainty can also be removed by the experimental data on $A_2(x,Q^2)$.

Although in this paper we have confined ourselves to the target mass effects 
in the Bjorken sum rule, the similar analysis can be carried out for the
Ellis-Jaffe sum rule. For the proton target, 
$\Delta\Gamma_{u.b.}^p(Q^2=2.5\mbox{GeV}^2)$=0.017, 0.016 and
0.0046 with typical errors of 10$\%$ for Analyses I, II and III, respectively.
Those for the neutron turn out to be 0.012 and 0.011 for Analyses I and II. 

Finally, we note that there exists the Burkhardt-Cottingham sum rule for 
$g_2(x,Q^2)$
\be
\int_0^1 dx g_2(x,Q^2)=0,
\ee
which is not only protected from QCD radiative 
corrections \cite{KMSU,ALNR,KMSU2} 
but also free from target mass effects \cite{MU}.

We hope that future experiments at CERN, SLAC and DESY will provide us with 
data on $A_1$ possessing higher statistics as well as the data on $A_2$ 
with high accuracy which will enable us to study $g_2$ structure functions 
and also target mass effects more in detail.


\newpage

\hspace{-2.5cm}
\epsfxsize=17cm
\epsffile{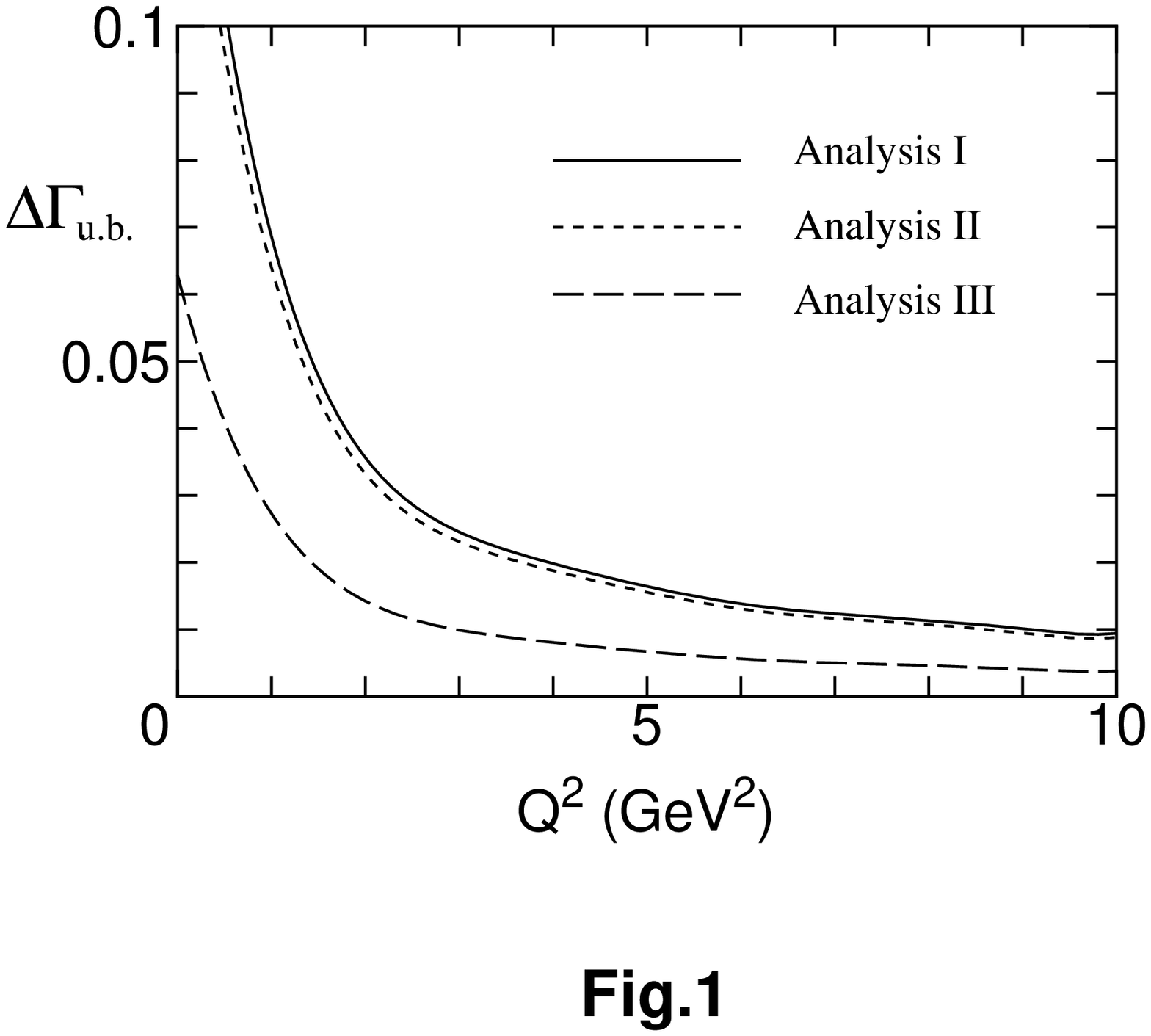}

\noindent
Fig.1 \quad The upper bound for the target mass correction $\Delta\Gamma$, 
$\Delta\Gamma_{u.b.}$, as a function of $Q^2$. The solid, short-dashed and 
long-dashed lines show the upper bounds for the analyses I, II and III, 
respectively.

\end{document}